\documentclass[10pt,aps,twocolumn,prl,showkeys,showpacs,superscriptaddress]{revtex4-2}

\usepackage[T1]{fontenc}
\usepackage[utf8]{inputenc}
\usepackage{graphicx}

\usepackage{dcolumn}
\usepackage{bm}
\usepackage{times}
\usepackage{graphics}
\usepackage{color}
\usepackage{soul}
\usepackage{setspace}
\usepackage{amsmath,amssymb}
\usepackage{ragged2e}
\usepackage{subfigure} 
\usepackage{natbib}
\usepackage{comment}
\usepackage{dirtytalk}
\usepackage[dvipsnames]{xcolor}
\usepackage[colorlinks=true,linkcolor=blue,citecolor=blue,urlcolor=blue]{hyperref}

\begin{document}

\preprint{APS/123-QED}

\title{
Unveiling Intrinsic Triplet Superconductivity in Noncentrosymmetric NbRe through Inverse Spin-Valve Effects
}

\author{F. Colangelo}
\affiliation{Dipartimento di Fisica ``E.R. Caianiello'', Universit\`{a} degli Studi di Salerno, I-84084 Fisciano (Sa), Italy}
\affiliation{CNR-SPIN, c/o Universit\`{a} degli Studi di Salerno, I-84084 Fisciano (Sa), Italy}

\author{M. Modestino}
\affiliation{Dipartimento di Fisica ``E.R. Caianiello'', Universit\`{a} degli Studi di Salerno, I-84084 Fisciano (Sa), Italy}

\author{F. Avitabile}
\affiliation{Dipartimento di Fisica ``E.R. Caianiello'', Universit\`{a} degli Studi di Salerno, I-84084 Fisciano (Sa), Italy}

\author{A. Galluzzi}
\affiliation{Dipartimento di Fisica ``E.R. Caianiello'', Universit\`{a} degli Studi di Salerno, I-84084 Fisciano (Sa), Italy}
\affiliation{CNR-SPIN, c/o Universit\`{a} degli Studi di Salerno, I-84084 Fisciano (Sa), Italy}

\author{Z. Makhdoumi~Kakhaki}
\affiliation{Dipartimento di Fisica ``E.R. Caianiello'', Universit\`{a} degli Studi di Salerno, I-84084 Fisciano (Sa), Italy}
\affiliation{CNR-SPIN, c/o Universit\`{a} degli Studi di Salerno, I-84084 Fisciano (Sa), Italy}

\author{A. Kumar}
\affiliation{Dipartimento di Fisica ``E.R. Caianiello'', Universit\`{a} degli Studi di Salerno, I-84084 Fisciano (Sa), Italy}
\affiliation{CNR-SPIN, c/o Universit\`{a} degli Studi di Salerno, I-84084 Fisciano (Sa), Italy}

\author{J. Linder}
\affiliation{Center for Quantum Spintronics, Department of Physics, Norwegian University of Science and Technology, NO-7491 Trondheim, Norway}

\author{M. Polichetti}
\affiliation{Dipartimento di Fisica ``E.R. Caianiello'', Universit\`{a} degli Studi di Salerno, I-84084 Fisciano (Sa), Italy}
\affiliation{CNR-SPIN, c/o Universit\`{a} degli Studi di Salerno, I-84084 Fisciano (Sa), Italy}
\affiliation{Centro NANO\_MATES, Universit\`{a} degli Studi di Salerno, I-84084 Fisciano (Sa), Italy}

\author{C. Attanasio}
\affiliation{Dipartimento di Fisica ``E.R. Caianiello'', Universit\`{a} degli Studi di Salerno, I-84084 Fisciano (Sa), Italy}
\affiliation{CNR-SPIN, c/o Universit\`{a} degli Studi di Salerno, I-84084 Fisciano (Sa), Italy}
\affiliation{Centro NANO\_MATES, Universit\`{a} degli Studi di Salerno, I-84084 Fisciano (Sa), Italy}

\author{C. Cirillo}
\affiliation{CNR-SPIN, c/o Universit\`{a} degli Studi di Salerno, I-84084 Fisciano (Sa), Italy}

\date{\today}

\begin{abstract}
NbRe is a non-centrosymmetric superconductor that has been proposed as a candidate for intrinsic spin-triplet pairing. However, a conclusive demonstration of triplet pairing in NbRe is yet to be found. To probe the presence of equal-spin triplet Cooper pairs, we fabricated Py/NbRe/Py trilayers capped with an antiferromagnetic layer. Magnetic and electrical measurements reveal an inverse spin-valve effect, which could indicate equal-spin triplet superconductivity. The minimal sample structure and the lack of \textit{ad hoc} engineered interfaces clearly associate our observation to intrinsic triplet correlations of NbRe. The availability of NbRe in thin-film form and the simplicity of the heterostructure highlight its potential as a scalable platform for superconducting spintronics.
\end{abstract}

\maketitle

Conventional $s$-wave superconductors, as described by Bardeen-Cooper-Schrieffer (BCS) theory~\cite{bcs1957}, are based on spin-singlet Cooper pairs formed by electrons with opposite spin and momentum. On the contrary, an intrinsic spin-triplet superconducting state,  pursued for a long time in different candidate materials such as UGe$_2$, URhGe, and Sr$_2$RuO$_4$~\cite{Aoki2001,Mackenzie2003,Kallin2012}, involves equal-spin pairing. However, despite decades of research, direct and conclusive evidence for intrinsic spin-triplet pairing still has to be found~\cite{Pustogow2019}. Due to this lack of intrinsic spin-triplet superconductors, many works focused on forcing conventional superconductor films into odd-frequency spin-triplet reservoirs by using \textit{ad hoc} engineered interfaces, complex multilayered structures and exotic material combinations~\cite{robinson2010,klose2012,banerjee2014}. The final goal is to provide a natural and scalable platform for dissipationless spin transport and superconducting spintronic devices, by enabling spin-polarized supercurrents and robust proximity effects in ferromagnetic environments~\cite{Bergeret2005,Eschrig2015,eschrig2011,linder2015}. 
In this respect, non-centrosymmetric superconductors (NCSs) have attracted considerable attention for the study of unconventional pairing states. The lack of inversion symmetry in their crystal structure lifts parity as a good quantum number, allowing the emergence of antisymmetric spin-orbit coupling (ASOC)~\cite{smidman2017}. When sufficiently strong, ASOC leads to a mixing of even-parity spin-singlet and odd-parity spin-triplet components in the superconducting order parameter~\cite{smidman2017}. Among the known NCSs, Nb$_{0.18}$Re$_{0.82}$ (NbRe) has emerged as a particularly interesting material~\cite{carla2015,carla2022,zahra2024}. Most candidate systems for intrinsic triplet superconductivity are complex and strongly correlated (e.g., heavy-fermion compounds), whereas NbRe is a chemically simple alloy. It crystallizes in the $\alpha$-Mn structure and exhibits accessible superconducting properties, with a bulk critical temperature of about 9~K ~\cite{carla2015}. In contrast to many proposed spin-triplet superconductors, it can be deposited as a thin film~\cite{carla2022,zahra2024}, a key requirement for device applications. Notably, point-contact spectroscopy and specific heat measurements have revealed the presence of two superconducting gaps in NbRe~\cite{carla2015} and, in addition, muon-spin rotation and relaxation studies showed time-reversal symmetry breaking~\cite{Shang2018}. Both findings were further supported by magnetoresistance and transport studies, again suggesting possible singlet-triplet admixture~\cite{sundar2019}. More recently, ferromagnetic resonance experiments on NbRe/Co/NbRe trilayers have pointed toward the presence of equal-spin triplet correlations through spin pumping signatures~\cite{carla2023}. However, despite these converging indications, a conclusive demonstration of triplet pairing in NbRe remains elusive.

\begin{figure}[!ht]
	\centering
	{\includegraphics[width=.52\textwidth]{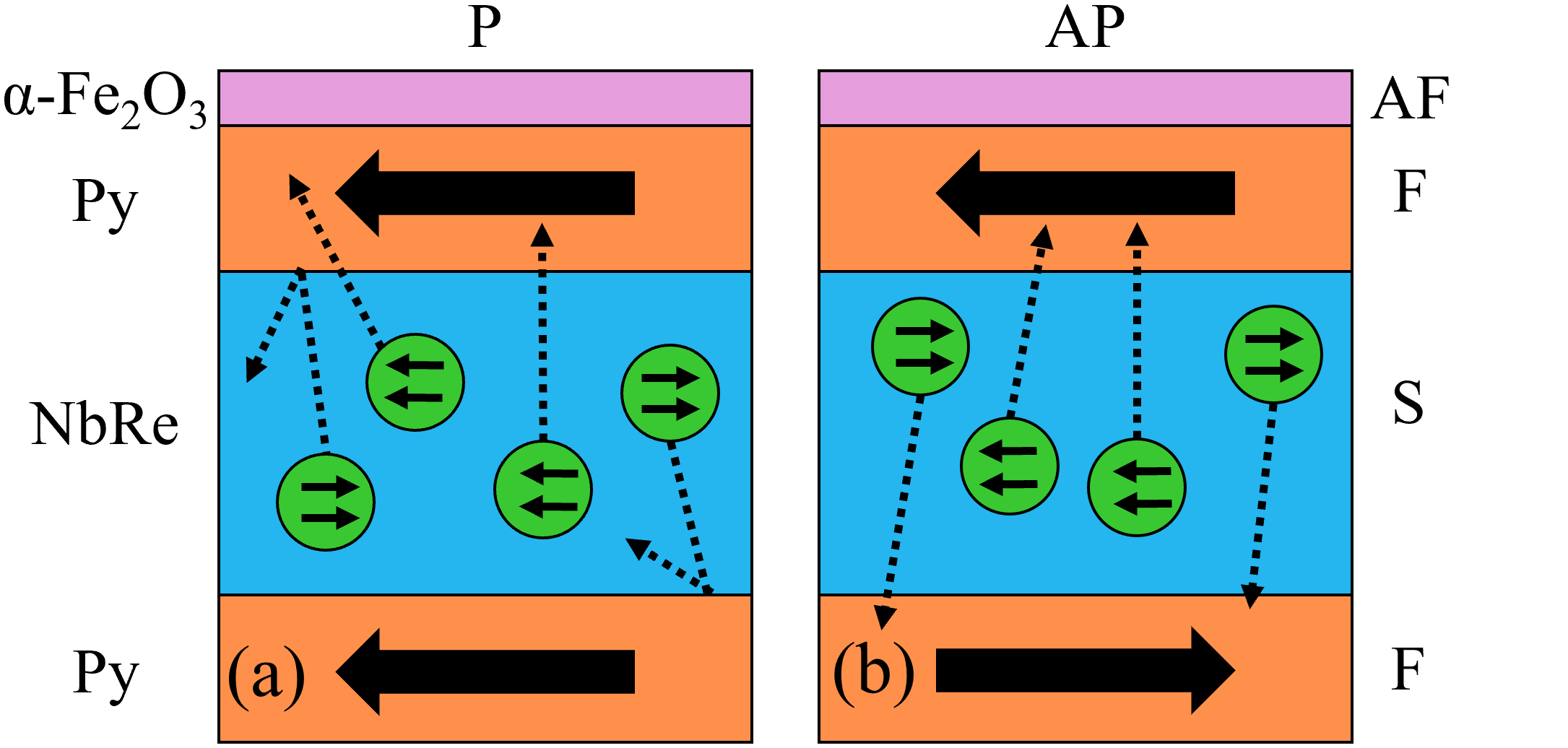}}
        \caption{Structure of the Py/NbRe/Py/$\alpha$-Fe$_2$O$_3$ (F/S/F/AF) SV device in both (a) P and (b) AP alignment of the two Py layer magnetizations. A representation of equal-spin triplet Cooper pairs and their propagation into the F layers is also depicted.}
        \label{SV_sample}
\end{figure}

\begin{figure}[!ht]
	\centering
	{\includegraphics[width=.48\textwidth]{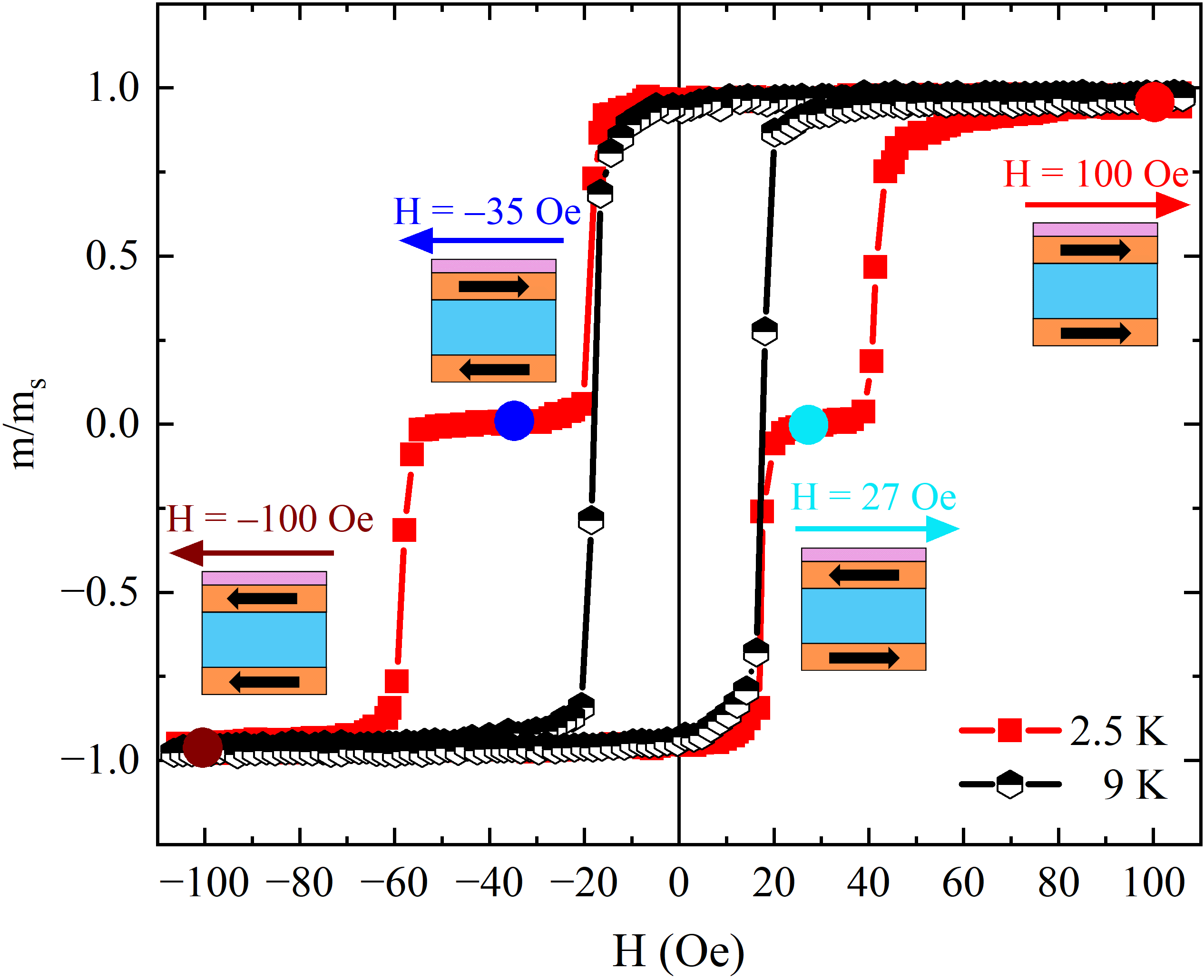}}
        \caption{
       Normalized magnetic moment $m/m_s$ of NbRe-based SV as a function of the magnetic field measured at $T=2.5$ and 9~K. The AP and P magnetic moments alignment of the two Py layers are schematically reported at the fields where the transport properties were performed. The colored arrows represent the orientation of magnetic field.}
        \label{Fig_Magn_NbRe}
\end{figure}

Taking advantage of the availability of NbRe in thin film form, we fabricated superconducting spin valve devices (SVs) based on NbRe and Ni$_{0.80}$Fe$_{0.20}$ (permalloy, Py) thin films to investigate the nature of its superconducting pairing. A SV consists of a superconductor layer (S) sandwiched between two ferromagnets (F) with independently controllable magnetizations~\cite{tagirov1999}. In such multilayers, the relative alignment of the ferromagnets---parallel (P) or antiparallel (AP)---modulates the effective exchange field acting on the superconductor. For singlet pairing, the AP configuration reduces pair-breaking and {typically} leads to an enhanced critical temperature $T_c$ compared to the P case~\cite{tagirov1999, buzdin1999, gu2002}, {although exceptions exist} \cite{rusanov_prb_06, steinar_prb_06, singh_prb_07, leksin_jetp_09, fominov_jetp_10}, {including for a different geometry (F/F/S) than considered here} \cite{leksin_prl_12, leksin_prb_16}. In contrast, for equal-spin triplet correlations, which align with the local magnetization, the AP configuration {allows the ferromagnets to respectively be proximitized primarily by Cooper pairs of opposite spin projections}, enhancing pair breaking and suppressing the $T_c$~\cite{banerjee2014}. {A higher $T_c$ in the P alignment is known as an} inverse spin-valve effect. Here, our aim is to study NbRe-based SV as a probing tool to investigate the nature of the superconducting pairing in this material.

Py/NbRe/Py trilayers, the core of the SVs, were deposited via ultra-high vacuum DC magnetron sputtering with base pressure in the low $10^{-8}$~mbar range on SiO$_2$ substrates. Stoichiometric Ni$_{0.80}$Fe$_{0.20}$ and Nb$_{0.18}$Re$_{0.82}$ targets were used. During deposition, the Ar pressure was kept at $P_{Ar}=6.5$~$\mu$bar via a mass flow controller, while sputtering powers of $W=150$~W and $W=350$~W were used for Py and NbRe, respectively. The deposition rates were monitored using a quartz crystal microbalance calibrated with a Bruker DektakXT profiler. Due to natural oxidation, a stable antiferromagnetic (AF) $\alpha$-Fe$_2$O$_3$ (hematite) layer formed on the top Py layer~\cite{Muan1958,Carla2013}, resulting in a final Py/NbRe/Py/$\alpha$-Fe$_2$O$_3$ structure (F/S/F/AF). The role of this AF layers is to pin the magnetization of the underlying Py, enabling a well-defined AP configuration. Moreover, it helps to suppress possible stray field-related artifacts discussed previously~\cite{Steiner2006,machiel2010}. It is worth noting that the native thermal oxide, spontaneously formed on the upper Py layer avoids the need for synthetic pinning or complex interface engineering, simplifying device realization. The structures, with thicknesses $d_{\mathrm{Py}}=12$~nm and $d_{\mathrm{NbRe}}=20$~nm, were deposited onto both unpatterned and patterned SiO$_2$ substrates, the latter featuring channel lengths and widths of $1500$~$\mu$m and $100$~$\mu$m, respectively. The choice of the Py thickness is expected to combine an efficient exchange-bias pinning by the hematite layer~\cite{Carla2013} and negligible effects from a potential dead-layer typically observed in ultrathin ferromagnetic films. {Notably, because the two Py layers are identical, our devices are protected against sign oscillations of the spin-valve effect. Therefore, any observed inverse effect is not consistent with predictions for a  conventional s-wave superconductor}~\cite{mironov2014}. Moreover, this choice ensures that the system operates outside the strong $T_c$ suppression regime associated with the $T_c(d_F)$ oscillatory behavior of F/S heterostructures~\cite{Buzdin2005}. Indeed, $d_{\mathrm{NbRe}}$ was also selected to ensure that the $T_c$ of the SV is in the operational range of our cryogenic setup. A representation of the samples is depicted in Fig.\hyperref[SV_sample]{\ref{SV_sample}}, where panels~\hyperref[SV_sample]{(a)} and \hyperref[SV_sample]{(b)} illustrate the P and AP configurations, respectively, with the magnetization directions of the two Py layers indicated by the black arrows. The scheme also highlights how, in the AP configuration, the equal-spin triplet pairs of both orientations can leak into the F layers, enhancing proximity-induced pair breaking and suppressing $T_c$, contrary to the behavior expected for a conventional SV. On the other hand, in P configuration, the leakage of Cooper pairs with opposite orientations with respect to the F layers is suppressed. These mechanisms will be further discussed in the following. A control SV device with the same structure was also fabricated by using a singlet superconductor to verify the validity of the experiment. For this purpose a Nb layer 25-nm-thick was embedded in the Py/Nb/Py/$\alpha$-Fe$_2$O$_3$ structure ($d_{\mathrm{Py}}=12$~nm). If the expected conventional behavior with $T_{c}^{P}$ < $T_{c}^{AP}$ is observed, any unconventional effects seen in the triplet-based device can be confidently attributed to the nature of the superconducting pairing rather than to spurious effects coming from the other layers. Nb was deposited at $P_{Ar}=3.5$~$\mu$bar and $W=350$~W, while Py was deposited using the same recipes as described above.

A detailed magnetic characterization by using a Vibrating Sample Magnetometer (VSM) option of a PPMS by Quantum Design was performed to investigate the behavior of the SVs deposited onto the unpatterned substrates. These measurements are crucial to test the reliably switching of the magnetic layers between the P and AP configurations, since the device must exhibit well-defined and reproducible magnetic states. Being the SV realized with soft ferromagnetic layers, such as Py, before every magnetic measurement, the residual field in the superconducting magnet was reduced as in~\cite{Modestino2023}. The dependencies of the magnetic moment on the temperature, $m(T)$, and on the DC magnetic field, $m(H)$, were studied. The $m(T)$, reported in the Supplemental Material (SM)~\cite{SM}, shows a transition at $T=9$~K, which can be associated to the blocking temperature ($T_B$) of the antiferromagnetic $\alpha$-Fe$_2$O$_3$ layer~\cite{Carla2013}. This native oxide layer pins the adjacent Py layer through exchange-bias interaction for $T<T_B$, while this effect is absent for $T>T_B$~\cite{Zaag2000,KUMAR2011802,Blachowicz2021}, when both the F layers are free to reverse.
This behavior can be observed in Fig.~\hyperref[Fig_Magn_NbRe]{\ref{Fig_Magn_NbRe}}, where the field dependence of the magnetic moment, normalized to its saturation value ($m_s$), is plotted. The curves were acquired at $T\leq T_B$ in the Field Cooling (FC) protocol, namely cooling the device from $T=20~$ K (i.e., above $T_B$) to the target temperature in a field of $H=3000$ Oe. Measurements were performed at different temperatures. The obtained $m(H)$ curves are comparable to the ones reported in literature for SV multilayers ~\cite{Perera1999,gu2002} and can be explained by considering the signal superposition of the two Py layers as illustrated in Fig.~\hyperref[SV_sample]{\ref{SV_sample}}. At large positive field, the Py layers are both magnetized in the field direction [P case in Fig.~\hyperref[SV_sample]{\ref{SV_sample}}(a)]. By reducing $H$ down to about $-20$~Oe, the magnetization of the free bottom Py layer switches in the field direction, while the upper F layer is pinned by the $\alpha$-Fe$_2$O$_3$ layer in the opposite direction due to the exchange interaction [AP case, in Fig.~\hyperref[SV_sample]{\ref{SV_sample}}(b)]. In this configuration, the magnetic signal is almost zero, as a further indication of the antiparallel alignment of the two Py layers. When further reducing $H$ to about $- 55$~Oe, the biased coercive field of the upper-Py is reached, and its magnetic moment also switches. Then, around $-65$~Oe, the P configuration is restored. By varying $H$ from -100~Oe up to +100~Oe, the antiparalled state is reached at the positive coercive field of bottom-Py (see the scheme at H~=~27~Oe in Fig.~\hyperref[Fig_Magn_NbRe]{\ref{Fig_Magn_NbRe}}) while the P state is finally restored when the exchange-biased positive coercive field of the upper-Py is overcome ($H=100$~Oe). Since the exchange-bias effect disappears at $T_B$, the $m/m_S(H)$ cycle measured at $T=9$ K is centered at the origin and shows the same coercivity as the free Py layer. This confirms that the transition observed in the $m(T)$ curve is related to the AF blocking. The effect of the temperature, with a progressive smearing of the $m/m_S(H)$ curves near $T_B$, is studied in more detail in the SM.

\begin{figure}[!ht]
	\centering
	{\includegraphics[width=.496\textwidth]{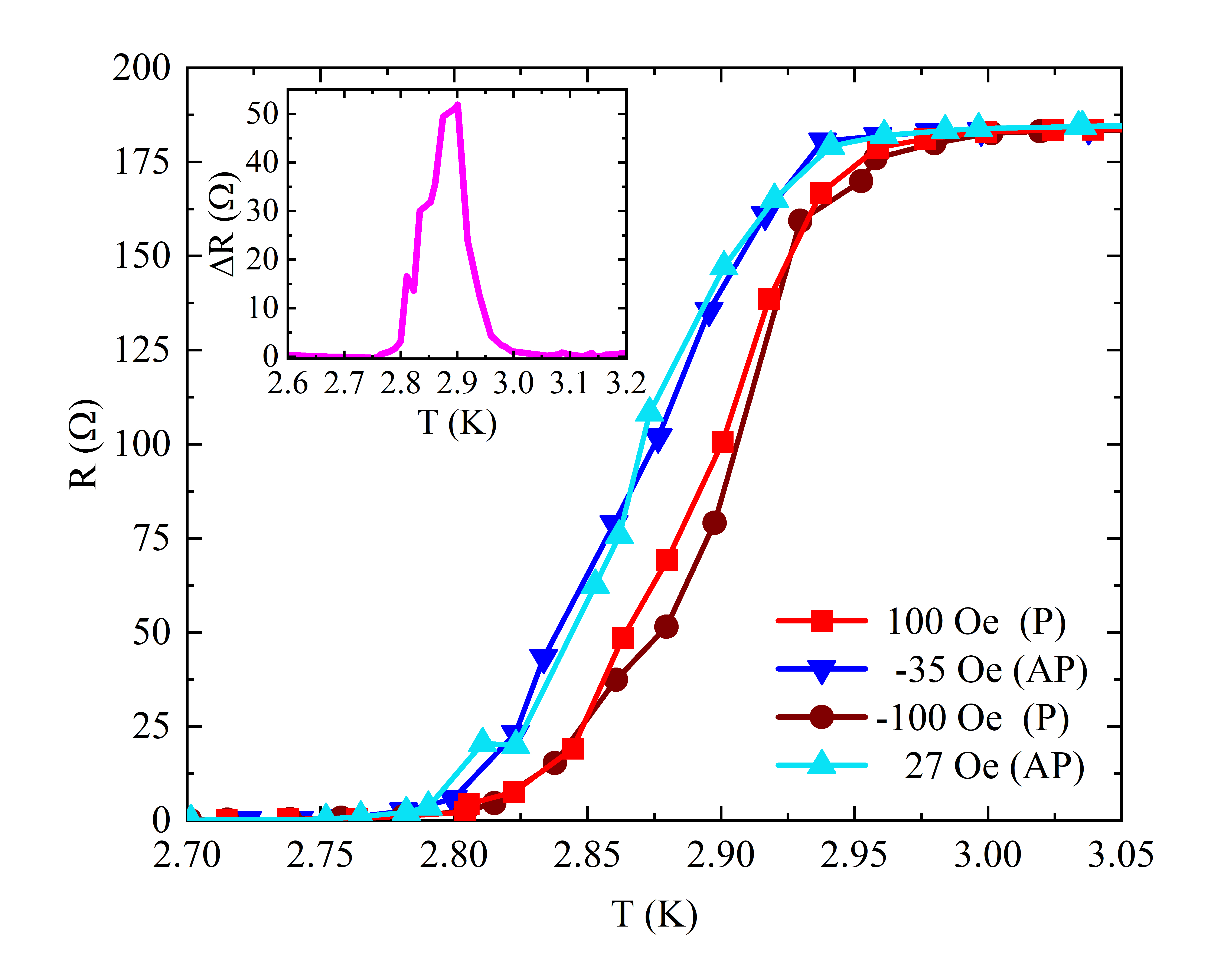}}
         \caption{Superconducting resistive transitions of NbRe-based SV measured in both P and AP magnetic configurations. The magnetic field values and the associated SV magnetic configuration (P or AP) are indicated in the legends. The difference between the averaged AP and P transitions is shown in the inset. See text for details.}
	\label{NbRe_SV_Tc}
\end{figure}

\begin{figure*}
    \centering
    {\includegraphics[width=1.01\textwidth]{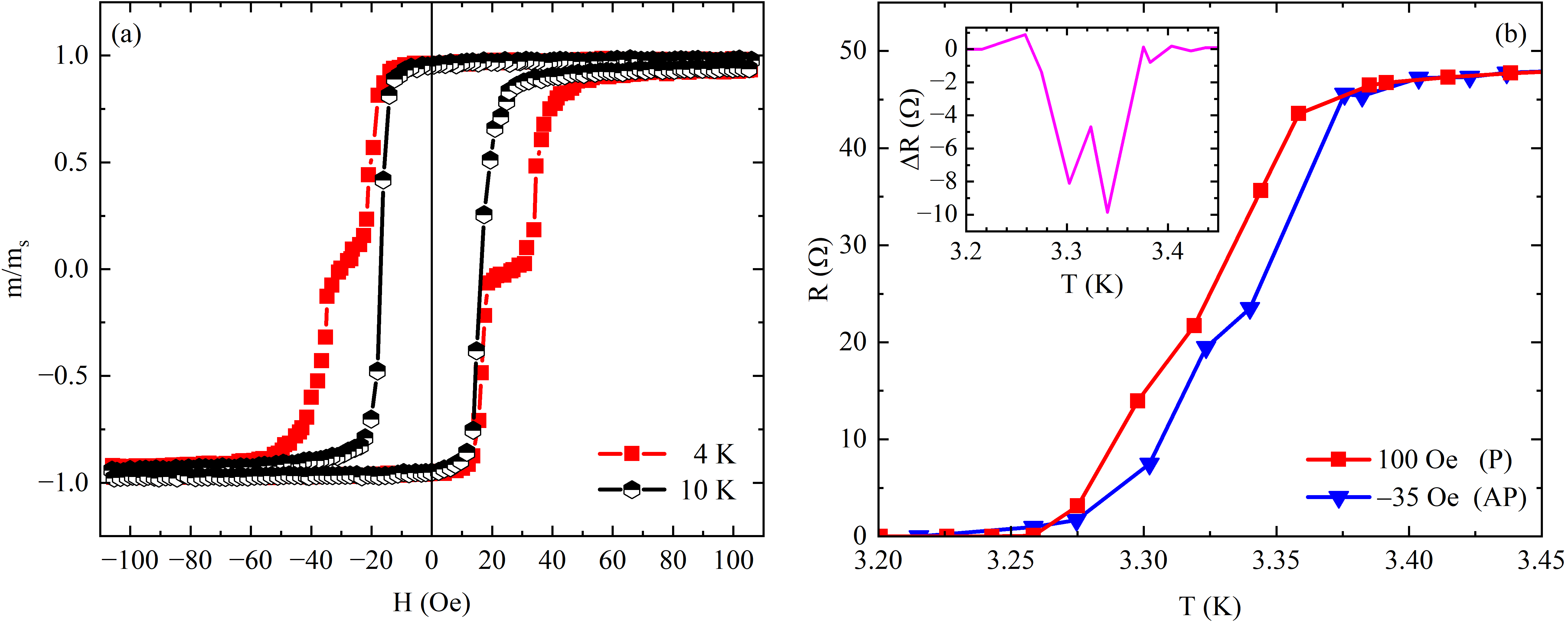}}    
    \caption{(a) Normalized magnetic moment $m/m_s$ of Nb-based SV as a function of the magnetic field measured at $T=4$ and 9~K. The AP and P magnetic moments alignment of the two Py layers can be observed in the curve performed at $T=4$~K. (b) Superconducting resistive transitions of Nb-based SV measured in both P and AP magnetic configurations. The magnetic field values and the associated P and AP configurations are reported in the legends. The difference between the averaged AP and P transitions is reported in the inset.}
    \label{Fig_Nb}
\end{figure*}

The spin-valve effect was investigated on devices deposited onto the patterned substrates using a Cryogen-Free High Field Measurement System (7~T) by CRYOGENIC Ltd. Electrical transport measurements were performed at magnetic field values selected from the preliminary $m(H)$ analysis [see Fig.~\hyperref[Fig_Magn_NbRe]{\ref{Fig_Magn_NbRe}(b)}]. To ensure accurate field calibration, the superconducting magnet was referenced with a Hall probe, and a Hall sensor was employed during the measurements. Resistance was measured in standard four-wire configuration using a Keithley~6121 current source and a Keithley~2182 nanovoltmeter operating in Delta mode, with a bias current of 10~$\mu$A.

Superconducting resistive transition measurements, $R(T)$, were performed in both P and AP configurations. The SV was cooled from $T=20$~K to $T=5$~K, under an in-plane magnetic field of $H=1000$~Oe to implement the FC protocol. The magnetic field was then set to $H=100$~Oe, $-35$~Oe, $-100$~Oe, and $27$~Oe, retracing the P and AP configurations of the $m(H)$ cycle in Fig.~\hyperref[Fig_Magn_NbRe]{\ref{Fig_Magn_NbRe}(b)}. $T_c$ was measured at each field, ensuring the temperature never exceeded $T_B$. The results are shown in Fig.~\hyperref[NbRe_SV_Tc]{\ref{NbRe_SV_Tc}}, where the superconducting transitions obtained in the P configuration are reported in red ($100$~Oe) and brown ($-100$~Oe), while the AP transitions in blue ($-35$~Oe) and cyan ($27$~Oe). The expected $T_c$ of a single 20-nm NbRe film is about $6.8$~K, with coherence length $\xi \sim 4.8$~nm~\cite{carla2016,carla2022}. However, the proximity effect from the adjacent Py layers leads to a significant suppression of the transition temperature. Defining $T_c$ at the $50$\% of the transition, the results clearly show that $T_c^{AP}<T_c^{P}$. To better quantify the superconducting transition in the two magnetic configurations, we averaged the resistance curves obtained in the P and AP states. Specifically, the two measurements taken in the P (AP) configuration were averaged to obtain a representative P (AP) curve. The $T_c$ evaluated from these curves were $T_c^{\mathrm{P}} = 2.90$~K and $T_c^{\mathrm{AP}} = 2.87$~K, revealing a difference of approximately 30~mK and thus confirming the inverse spin-valve effect. Furthermore, in the inset of Fig.~\hyperref[NbRe_SV_Tc]{\ref{NbRe_SV_Tc}} the difference between the average AP and P transition is reported. To evaluate the resistance variation between the two configurations across the transition, the average AP curve was subtracted from the average P curve. The result, shown in the inset of Fig.~\ref{NbRe_SV_Tc}, highlights a resistance difference exceeding 50~$\Omega$ near $T = 2.90$~K. Notably, the full switching between magnetic configurations, and hence between distinct superconducting states, was achieved with in-plane fields of only a few tens of oersteds, confirming suitability for cryogenic applications. 

The suppression of $T_c$ in the AP alignment of the NbRe-based SV contrasts with the conventional singlet proximity effect, which predicts a higher $T_c$ in the AP state due to partial cancellation of the exchange field~\cite{Buzdin2005}, as observed in Nb-based devices~\cite{gu2002}. Instead, the inverse spin-valve effect seen in NbRe strongly implies the involvement of equal-spin triplet Cooper pairs. These pairs are robust against exchange-induced dephasing and can propagate over long distances into ferromagnets. The modulation of $T_c$ with magnetic alignment reflects a spin-selective absorption of these triplet components at the S/F interfaces~\cite{romano2024}, as previously discussed by Banerjee et al.~\cite{banerjee2014}, where triplet pairs were externally induced via a spin-mixer layer. In our case, although no such spin-mixer is present, the mechanism is analogous: ferromagnetic layers with defined magnetization preferentially absorb one triplet pair with the same orientation while reflecting the other. When the F layers are aligned (P configuration), this selectivity limits leakage to a single spin species. In the AP configuration, both components can propagate into the ferromagnets, resulting in larger depletion of superconducting correlations and suppression of $T_c$. Crucially, here the observed triplet features arise without engineered spin-active interfaces or spin-mixing layers and the simplicity of the structure examined in the experiment suggests that the observed result is unlikely to be due to artifacts. Therefore, the presence of intrinsic equal-spin correlations can reasonably originate from the lack of inversion symmetry in the NbRe crystal structure.

In order to rule out possible mixing and spin-triplet generation mechanisms due to the interfaces, the experiments were repeated on the Nb-based SVs. In fact, not only is Nb a well-known singlet s-wave superconductor, but Nb-based SVs have already showed conventional spin-valve effect~\cite{gu2002}. The magnetic characterization of the Nb-based SV was performed following the same procedure described above. The $m(H)$ curves in Fig.~\hyperref[Fig_Nb]{\ref{Fig_Nb}(a)} confirm the presence of AP and P configurations, as observed for the Py/NbRe/Py/$\alpha$-Fe$_2$O$_3$ multilayer. As further described in the SM, the $m(T)$ and the $m(H)$ curves also show hysteresis loops associated with the superconducting behavior of Nb. The superconducting transition of the Nb-based SV are reported in Fig.~\hyperref[Fig_Nb]{\ref{Fig_Nb}(b)}. As expected for conventional $s$-wave superconductivity, the AP configuration yields a larger $T_c$ due to partial compensation of the exchange fields at the two F/S interfaces. In the inset of Fig.~\hyperref[Fig_Nb]{\ref{Fig_Nb}(b)}, the difference between the P and AP curves is reported, and the observed $\Delta T_c \approx -20$~mK confirms the presence of standard spin-singlet proximity behavior, in agreement with Ref.~\cite{gu2002}. {Here, the minus sign of $\Delta T_c$ indicates that this values is referred to the standard spin valve effect.} This result further ensures that the inverse spin-valve effect of NbRe-based SV is not due to structural and magnetic artifacts, such as stray fields and domain walls~\cite{machiel2010,Steiner2006}. We also comment on the possible influence of inverse crossed Andreev reflection (iCAR) on $T_c$. In this process, the electrons in a Cooper pair are split non-locally into one ferromagnet each, which enhances $T_c$ in the P configuration. We believe that this mechanism can be ruled out in the present measurements for the following reason. The competition between the pair-breaking due to induced magnetization, favoring an AP alignment, and the pair-breaking due to iCAR, favoring a P alignment, depends on the interface transparency of the junction. The NbRe/Py and Nb/Py interfaces used in our study have very similar transparency, as shown (see SM) by the fact that their normalized $T_c$-curves essentially coincide with each other. Taking further into account that NbRe and Nb also have similar coherence lengths~\cite{carla2016,NJP2017} and intrinsic critical temperatures~\cite{carla2016,Carla2005}, if iCAR were the dominant pair-breaking mechanism leading to higher $T_c$ in the P alignment, this should also have been seen for the Py/Nb/Py measurements. However, as discussed above, this is not the case, and therefore iCAR can likely be ruled out as the mechanism leading to higher $T_c$ in the P alignment.

In conclusion, we fabricated Py/NbRe/Py/$\alpha$-Fe$_2$O$_3$ spin-valve devices and investigated their magnetization and transport properties. The magnetic characterization showed two achievable magnetic configurations of the Py layer, with P or AP magnetization. The analysis of the superconducting resistive transitions revealed a reproducible suppression of $T_c$ in the AP configuration relative to the P state, providing direct evidence for spin-selective leakage of intrinsic equal-spin triplet pairs in noncentrosymmetric NbRe. The ability to modulate superconductivity with small magnetic fields highlights the potential of these devices as superconducting switches. It is also fundamental to note that the samples were realized in thin film form, a crucial aspect in both device fabrication and fundamental research experiments.
Furthermore, their structurally minimal and self-assembled architecture—comprising common metallic layers and a naturally oxidized cap—offers clear advantages for scalable integration in superconducting spintronic technologies.

\bibliographystyle{apsrev4-2}
\bibliography{bibliography}

\end{document}